\documentclass[10pt, reqno]{amsart}
\usepackage{float}
\usepackage{graphicx}
\usepackage{booktabs}
\usepackage{orcidlink}
\usepackage{caption}
\usepackage{multirow}
\usepackage{indentfirst,csquotes}
\usepackage{amssymb,amsthm,amsmath}
\usepackage{xcolor,paralist,hyperref,fancyhdr,etoolbox}
\usepackage{algorithm}
\usepackage{algpseudocode}

\hypersetup{ colorlinks=true, linkcolor=black, filecolor=black, urlcolor=black, citecolor=black }

\makeatletter
\def\section{\@startsection{section}{1}%
  \z@{.7\linespacing\@plus\linespacing}{.5\linespacing}%
  {\normalfont\bfseries}}

\def\subsection{\@startsection{subsection}{2}%
  \z@{.5\linespacing\@plus.7\linespacing}{.3\linespacing}%
  {\normalfont\bfseries}}

\let\oldsection\section

\renewcommand{\section}{%
  \@ifstar{\section@star}{\section@nostar}%
}

\def\section@star#1{%
  \oldsection*{\MakeUppercase{#1}}%
}

\def\section@nostar#1{%
  \oldsection{\MakeUppercase{#1}}%
}

\renewcommand{\@seccntformat}[1]{%
  \csname the#1\endcsname\quad$|$\quad}
\makeatother

\begin{document}

% Manual title formatting
\begin{center}
{\Large\bfseries Virtual EMI Receiver Generation (VERGE) for Improved EMI Removal in MRI}

\vspace{1em}

Alexander Mertens$^{1,2}$\,\orcidlink{0000-0001-5172-0444}, 
Diego Martinez$^{1,2}$\,\orcidlink{0009-0004-4272-7782},
Amgad Louka$^{1}$\,\orcidlink{0009-0003-5595-5381}, 
Ying Yang$^{1,2}$\,\orcidlink{0009-0002-8804-566X}, 
Chad Harris$^{4}$\,\orcidlink{0009-0001-2004-468X}, 
Ian Connell$^{1,2,3,5}$\,\orcidlink{0000-0002-7722-3603}

\vspace{0.5em}

{\small
$^1$Biomedical Engineering, University Health Network, Toronto, ON, CA\\
$^2$Medical Biophysics, Temerty Faculty of Medicine, University of Toronto, Toronto, ON, CA\\
$^3$CenteR for Advancing Neurotechnical Innovation to Applications, University Health Network, Toronto, ON, CA\\
$^4$Research and Development, Synaptive Medical, Toronto, ON, CA\\
$^5$The KITE Institute, University Health Network, Toronto, ON, CA\\
}

\vspace{0.5em}

{\small Correspondence: Ian.Connell@uhn.ca}

\vspace{0.5em}

{\small\today}
\end{center}

\vspace{1em}

\noindent{\small\textit{Funding:} This work was partially supported by Ontario Research Fund Research Excellence program.}

\vspace{0.5em}

\noindent{\small\textit{Keywords:} MRI, Electromagnetic Interference, STRIDE, MEENO, low-field MRI}

\begin{abstract}
\section{Purpose} MRI is not designed to operate in the presence of EMI emitting electronic devices. To accomodate this, we propose VERGE which improves on previous EMI removal methods and can operate without external EMI sensors.
\section{Methods and Theory} VERGE estimates a reference coil non-linearly from all imaging and non-imaging coils by training a small per-scan convolutional neural net to generate a virtual coil. The virtual coil is then used with STRIDE, a reference-based EMI removal method.
\section{Results} For single-source EMI contamination, VERGE performed similarly to the reference method. For multi-source dynamic EMI contamination, VERGE outperformed the comparison methods. Notably, VERGE with zero physical EMI sensors outperformed the comparison methods even with two physical EMI sensors

\section{Conclusion} A non-linearly estimated virtual EMI coil improves reference-coil-based EMI removal under dynamic multi-source EMI, and enables effective EMI removal without physical EMI sensors or training data.
\end{abstract}

\bigskip
\section{Introduction}\label{sec1}

\subsection{Motivation}

Magnetic resonance imaging (MRI) provides excellent soft tissue contrast, but its high sensitivity leaves it susceptible to electromagnetic interference (EMI). This occurs when electronic devices near the scanner emit electromagnetic signals or produce harmonics near the Larmor frequency. These signals are measured by the imaging coils and severely degrade image quality. MR scanners are typically built inside radiofrequency (RF) shielded rooms to prevent any contamination from external EMI emitting devices. 

A growing number of emerging applications, like interventional MRI and acute imaging, require that MR scanners operate in the presence of devices that can severely degrade image quality if not meticulously designed to operate in the MR environment \cite{Iordachita, jellema2023, wach2025, Marincola2018MRI, Madewell2019Neurosurgical, rogers2021, pawlik2024, KUMAR2024107662, puhrwesterheide2022cost, orlando}. This is especially challenging in environments with generic electronic equipment like the operating room, emergency department, or intensive care unit spaces where devices are not designed with MR in mind, or when scanners are low field and devices can be brought closer without becoming a safety risk \cite{kuoy2022poc_mri, panther2019dedicated}.

\subsection{Existing Methods}

Passive shielding mitigates EMI without external sensors by enclosing only the scanner in a conductive shield rather than the entire room. Unless the patient is also shielded, which is generally impractical, the patient will act as a waveguide and the images will still be corrupted by EMI \cite{OReilly2021, Cooley2021}.

Algorithmic EMI removal has a long history outside of MRI in fields like geophysics, biomedical instrumentation, and surface NMR \cite{gamble1979, osti_7023612, spies1988, espy2015, walsh2008, muller-petke2020, ren2017, EDITER, letcher1989, STRIDE, wang2026meeno}. Unlike passive shielding, algorithmic methods use external sensors to measure EMI independently from the signal of interest. An EMI estimation and removal algorithm uses the reference sensor to clean the primary sensor.

External Dynamic Interference Estimation and Removal (EDITER) has become a standard for deterministic EMI removal in the field of MRI \cite{EDITER}. EDITER clusters acquired data into statistically stationary temporal groups, and estimates a linear transfer function to describe the relationship between the EMI sensing coils and the EMI corrupted primary imaging coils to clean the data. Smooth Total Variation Regularization for Interference Detection and Elimination (STRIDE) improves on EDITER by incorporating prior known MR image properties into the EDITER problem formulation \cite{STRIDE}.

More recent multi-channel methods also exploit correlations among the imaging coils themselves. Inter-Channel Correlation-Based EMI Noise Removal (ICER) combines EMI sensor measurements with inter-channel correlation among the receive coils \cite{huang2025icer}, and Multi-channel Electromagnetic Interference Elimination using Null Operations (MEENO) uses null operations across all receive and sensor channels to remove EMI, performing well even with few or no external sensors \cite{huang2025icer, wang2026meeno}.

Machine learning (ML) approaches also exist which typically use dedicated EMI characterization windows in each TR to acquire calibration data. Some methods use this calibration data to learn the relationship between EMI sensing coils and primary imaging coils to predict the EMI contaminating the primary coils, whereas other methods learn the EMI cleaned data directly. Deep Learning Direct MR Signal Prediction (Deep-DSP) can also be trained without a dedicated characterization window by treating the periphery of k-space as pure EMI \cite{ML1, deepdsp}. Deep-DSP$^2$ builds on this by operating in both the spectral and frequency domains, and by using a more sophisticated neural net for EMI removal \cite{hu2027deepdsp2}. The limitations of ML methods are that they are not designed to operate without physical EMI sensors, and that methods which directly predict the EMI-free signal require EMI-free MR training data \cite{deepdsp}.

\subsection{Proposed Approach}

In this work, we propose Virtual EMI Receiver Generation (VERGE), an EMI removal method that does not require physical EMI sensors, pulse sequence modification, or EMI-free training data. VERGE non-linearly estimates a single virtual EMI coil from all primary and EMI sensing coils using a small CNN optimized on each acquisition individually, and applies it with subsequent reference-coil based EMI removal methods. We evaluate VERGE on \textit{in vivo} data acquired under two single source and one dynamic multisource EMI scenario. For each scenario, we remove EMI with two physical EMI reference coils, and again with zero physical EMI reference coils. We compare VERGE to MEENO followed by STRIDE when two physical sensors are used, and just MEENO when zero physical sensors are used. We also compare VERGE to Deep-DSP and Deep-DSP$^2$ when two physical EMI sensors are used. To evaluate theses methods, we compute SNR, EMI removal percentage, and use visual comparison.

\section{Theory}\label{theory}

\subsection{VERGE Pipeline Overview}

VERGE generates a virtual reference coil from the primary imaging coils that can be used by any subsequent reference-coil EMI removal method. In this work, a good virtual coil will both reduce EMI in the background to near zero, and minimize correlation between the EMI-removed image and the EMI itself. The pipeline consists of four steps:

\begin{enumerate}
  \item \textbf{Initial EMI removal (Section~\ref{theory:meeno}).} MEENO is applied to the raw k-space to produce partially EMI-cleaned coil images. The result of this, $x_{\text{MEENO}}$, is used as an input for subsequent steps.
  \item \textbf{Background mask generation (Section~\ref{theory:mask}).} The virtual-coil loss in step 3 measures residual EMI only in the background, which requires a mask that excludes anatomy. To obtain it, a preliminary virtual noise coil is generated by a CNN trained with spatial total variation (TV) as its only objective. The resulting image is smooth enough that thresholding separates anatomy from background.
  \item \textbf{Virtual noise coil generation (Section~\ref{theory:vnc}).} A CNN of the same architecture is retrained to generate a new virtual coil with a different objective. The CNN will produce a virtual reference coil that results in minimal residual EMI in the background, and minimal correlation between the EMI that $x_{\text{MEENO}}$ was able to remove, and EMI-removed image with the virtual coil.
  
  \item \textbf{Final EMI removal (Section~\ref{theory:stride}).} Finally, STRIDE is used to remove EMI using the virtual coil learned in step 3 along with any additional EMI sensors.
\end{enumerate}

\subsection{Initial EMI removal with MEENO}\label{theory:meeno}

MEENO begins by constructing a block Hankel matrix from a sliding window of length \textit{w} along the readout direction of each phase encode line. At each window position, data from the C imaging coils and \textit{n} EMI sensors are concatenated into a column of the Hankel matrix, $\mathbf{H}$.

MEENO is calibrated using the peripheral $25\%$ of k-space phase encode lines. In this work, no separate calibration scan is used to aide reconstruction. Similar to above, a  Hankel matrix,  $\mathbf{H}_C$, is constructed using the calibration data and singular value decomposition is performed:

\begin{equation}
\mathbf{H}_C = \mathbf{U}_C  \mathbf{S}_C  \mathbf{V}_C^{*}
\end{equation}

Here, $\mathbf{S}_C$ is a diagonal matrix of singular values, $\mathbf{U}_C$ is the left singular
vector matrix, and $\mathbf{V}_C$ is the right singular vector matrix. The left singular vectors associated with the $r$ largest singular values approximately span the EMI subspace. These are excluded, and the remaining left singular vectors are stored as $\mathbf{U}_n$.

Then, the columns of $\mathbf{H}$ are projected onto the subspace spanned by the columns of $\mathbf{U}_n$:
\begin{equation}
\mathbf{H}' = \mathbf{U}_n  \mathbf{U}_n^{*}  \mathbf{H}
\end{equation}

The cleaned data, $x_{\text{MEENO}}$, are extracted by performing a de-Hankel operation on $\mathbf{H}'$.

\subsection{Background Mask Generation}\label{theory:mask}
Thresholding $x_{\text{MEENO}}$ directly results in a poor mask since $x_{\text{MEENO}}$ contains residual EMI. Therefore, a preliminary virtual coil-based EMI removal is performed that is designed only to distinguish between anatomy and air. A virtual coil, $V_\theta$, is generated by a CNN, $f_\theta$, from the acquired coil images, $X$, split into real and imaginary channels:

\begin{equation}
V_\theta = f_\theta\big([\,\mathrm{Re}\,X,\ \mathrm{Im}\,X\,]\big)
\end{equation}

Here, $\theta$ are the network weights. The network consists of four $3\times3$ convolutional layers with 64 channels, ReLU activations, and dilation factors of 1, 2, 4, and 1, followed by a $1\times1$ convolution to two output channels that form the real and imaginary parts of $V_\theta$. No pooling is used, so the virtual coil has the same resolution as the input. The network is trained independently for each acquisition and is not intended to generalize to other data. Instead, the convolutional architecture acts as an implicit spatial regularizer on $V_\theta$, similar to the deep image prior\cite{ulyanov2018dip}.
 
The network weights are chosen to minimize the following objective function:

\begin{equation}
\begin{gathered}
\theta_{\text{TV}} = \operatorname*{argmin}_{\theta}\
\frac{\mathrm{TV}\Big(\mathrm{SoS}\big(\mathrm{STRIDE}\big(x_{\text{MEENO}},\ [\,E,\ V_\theta\,]\big)\big)\Big)}
     {\mathrm{TV}\Big(\mathrm{SoS}\big(\mathrm{STRIDE}\big(x_{\text{MEENO}},\ E\big)\big)\Big)},
\\[6pt]
s_{\text{TV}} = \mathrm{SoS}\big(\mathrm{STRIDE}\big(x_{\text{MEENO}},\ [\,E,\ V_{\theta_{\text{TV}}}\,]\big)\big)
\end{gathered}
\end{equation}
 
Here, $E$ are the images from the $n$ physical EMI sensors ($E=\varnothing$ when $n=0$, in which case the denominator reduces to $\mathrm{TV}(\mathrm{SoS}(x_{\text{MEENO}}))$), $\mathrm{SoS}$ is the root sum-of-squares over coils, $\mathrm{TV}$ is the spatial total variation, and $s_{\text{TV}}$ is the resulting preliminary image. The denominator normalizes the objective so that it equals 1 when the virtual coil removes nothing.
 
The resulting $s_{\text{TV}}$ is thresholded, dilated by 10 voxels, and inverted to generate the air mask, $\Omega_{\text{air}}$.

\subsection{Virtual Noise Coil Generation}\label{theory:vnc}

The same CNN is retrained to generate a new virtual coil, $V_\theta$, with a different objective. The resulting virtual coil should produce minimal residual EMI in the background, and minimal correlation between the EMI already removed by MEENO and the EMI-removed image. Before providing the objective function, we define the following images:

\begin{align}
s_c &= \mathrm{SoS}\big(\mathrm{STRIDE}\big(x_{\text{MEENO}},\ [\,E,\ V_\theta\,]\big)\big)\\
s_{\text{M}} &= \mathrm{SoS}\big(\mathrm{STRIDE}\big(x_{\text{MEENO}},\ E\big)\big)\\
x_{\text{emi}} &= \mathrm{SoS}\big(x_{\text{raw}} - \mathrm{STRIDE}\big(x_{\text{MEENO}},\ E\big)\big)\\
s_{\text{raw}} &= \mathrm{SoS}\big(x_{\text{raw}}\big)
\end{align}

Here, $s_c$ is the image after EMI removal with the virtual coil, $s_{\text{M}}$ is the image without the virtual coil ($s_{\text{M}} = \mathrm{SoS}(x_{\text{MEENO}})$ when $n=0$), $x_{\text{raw}}$ are the
uncorrected coil images, $x_{\text{emi}}$ is the SoS image of the EMI already removed, and $s_{\text{raw}}$ is the uncorrected image.

The objective function consists of two terms:
\begin{align}
\mathcal{L}_1 &= \frac{\sum_{\Omega_{\text{air}}} s_c^2}{\sum_{\Omega_{\text{air}}} s_{\text{M}}^2}\\[4pt]
\mathcal{L}_2 &= \frac{\rho^2\big(x_{\text{emi}},\, s_c\big) - \rho^2\big(x_{\text{emi}},\, s_{\text{M}}\big)}
     {\rho^2\big(x_{\text{emi}},\, s_{\text{raw}}\big) - \rho^2\big(x_{\text{emi}},\, s_{\text{M}}\big)}
\end{align}

The denominators normalize $\mathcal{L}_1$ and $\mathcal{L}_2$ such that, without a virtual coil,
$\mathcal{L}_1 = 1$ and $\mathcal{L}_2 = 0$. Network weights are chosen to minimize their weighted sum:
\begin{equation}
\theta^{\star} = \operatorname*{argmin}_{\theta}\ \mathcal{L}_1 + \gamma\,\mathcal{L}_2
\end{equation}\label{eqn:obj}
Here, $\rho^2$ is the squared Pearson correlation and $\gamma$ is an empirically chosen weight.

\subsection{STRIDE Removal with the Virtual Noise Coil}\label{theory:stride}

After training, the final virtual coil, $V_{\theta^{\star}}$, is used together with any physical EMI
sensor images, $E$, as the reference set for a final STRIDE removal \cite{STRIDE}:
\begin{equation}
x_{\text{VERGE}} = \mathrm{STRIDE}\big(x_{\text{MEENO}},\ [\,E,\ V_{\theta^{\star}}\,]\big)
\end{equation}
Here, $x_{\text{VERGE}}$ are the final EMI-removed coil images.

Pseudocode for the algorithm can be found in algorithm \begin{algorithm}[!t]
\caption{VERGE}
\label{alg:verge}
\begin{algorithmic}[1]
    \Require k-space from $C$ imaging coils and $n$ EMI sensors, number of iterations $T$, weight $\gamma$
    \Ensure EMI-removed coil images $x_{\text{VERGE}}$
    \Statex
    \Statex \textbf{Initial EMI removal}
    \State Inverse Fourier transform the imaging coils to obtain $x_{\text{raw}}$ and the sensors to obtain $E$ ($E=\varnothing$ when no physical sensors are used)
    \State $x_{\text{MEENO}} \gets \mathrm{MEENO}(\text{k-space})$
    \State $s_{\text{M}} \gets \mathrm{SoS}\big(\mathrm{STRIDE}(x_{\text{MEENO}}, E)\big)$
    \State $x_{\text{emi}} \gets \mathrm{SoS}\big(x_{\text{raw}} - \mathrm{STRIDE}(x_{\text{MEENO}}, E)\big)$
    \State $s_{\text{raw}} \gets \mathrm{SoS}(x_{\text{raw}})$
    \Statex
    \Statex \textbf{Background mask generation}
    \State Initialize network weights $\theta$
    \For{$t \in \{1, \ldots, T\}$}
        \State $V_\theta \gets f_\theta\big([\,\mathrm{Re}\,X,\ \mathrm{Im}\,X\,]\big)$
        \State $s_c \gets \mathrm{SoS}\big(\mathrm{STRIDE}(x_{\text{MEENO}}, [\,E,\ V_\theta\,])\big)$
        \State Update $\theta$ to minimize $\mathrm{TV}(s_c)/\mathrm{TV}(s_{\text{M}})$
    \EndFor
    \State $s_{\text{TV}} \gets s_c$
    \State Threshold, dilate, and invert $s_{\text{TV}}$ to obtain $\Omega_{\text{air}}$
    \Statex
    \Statex \textbf{Virtual noise coil generation}
    \State Reinitialize network weights $\theta$
    \For{$t \in \{1, \ldots, T\}$}
        \State $V_\theta \gets f_\theta\big([\,\mathrm{Re}\,X,\ \mathrm{Im}\,X\,]\big)$
        \State $s_c \gets \mathrm{SoS}\big(\mathrm{STRIDE}(x_{\text{MEENO}}, [\,E,\ V_\theta\,])\big)$
        \State Compute $\mathcal{L}_1$ and $\mathcal{L}_2$ from $s_c$, $s_{\text{M}}$, $x_{\text{emi}}$, $s_{\text{raw}}$, and $\Omega_{\text{air}}$
        \State Update $\theta$ to minimize $\mathcal{L}_1 + \gamma\,\mathcal{L}_2$
    \EndFor
    \State $\theta^{\star} \gets \theta$
    \Statex
    \Statex \textbf{Final EMI removal}
    \State $x_{\text{VERGE}} \gets \mathrm{STRIDE}\big(x_{\text{MEENO}}, [\,E,\ V_{\theta^{\star}}\,]\big)$
\end{algorithmic}
\end{algorithm}

\section{Methods}

All experiments were performed on a head-only 0.5T scanner (Synaptive Medical, Toronto, Canada) equipped with a 16 channel head-coil and 2 custom-built EMI sensors (Figure \ref{sup:figure_s1}).

\subsection{EMI sensor Construction}

The in-house-constructed two EMI sensors were identical to those described in STRIDE \cite{STRIDE}: 2-loop solenoids of 16 AWG copper wire with a 25 mm diameter, each connected to a preamplifier (Synaptive Medical, Toronto, Canada) and placed orthogonally behind the head coil to cover the XY plane, representative of the transverse magnetization and previously demonstrated to be sufficient for full coverage of external EMI \cite{STRIDE}.

\subsection{Data Acquisition}

In-vivo brain data was collected on two human volunteers on two human volunteers with informed consent and approval of the University Health Network Research Ethics Board (REB\#20-5175), using a 2D gradient echo (GRE) pulse sequence with a flip angle of 22 degrees, a matrix size of 256x256, an FOV of 256-mm, 4-mm slice thickness, a TR of 50ms, and a TE of 7.238ms. Ten temporal phases were acquired for each of the following EMI scenarios: (1) baseline acquisition (no EMI), (2, Square) function generator operating at 21.28 MHz with amplitude modulation by a 10 kHz square wave at -50 dB, (3, Vital sign monitor) Welch Allyn (WA) vital sign monitor with pulse oximeter and thermometer placed near the scanner bore contacting the subject, (4, multi-source dynamic) a WA vital sign monitor as described in scenario 3 with a square wave noise generator as described in scenario 2 turned on during the 5th temporal phase and moved around the room randomly by hand for the remainder of the experiment.

\subsection{Data Reconstruction}

Each of the sets of 10 images were reconstructed individually in the following ways: VERGE with two physical EMI reference coils, VERGE with zero EMI reference coils, MEENO + STRIDE with two physical EMI reference coils, and MEENO with zero EMI reference coils. An additional image was computed by taking the average of all 10 reconstructed images. The parameters for both VERGE and MEENO were chosen to produce the best results for each method individually. Each implementation of STRIDE used $\Delta y = 7$ as defined in the original publication. For MEENO, a window length of $w=5$ and a cutoff of $r=15$ was used. The value of $\gamma$ in equation \ref{eqn:obj} was set to 0.3, which was determined through a grid search.

Deep-DSP and Deep-DSP$^2$ were both implemented as per their publicly available code and publications and were only applied to the datasets with two physical EMI sensors, since both methods require EMI sensor signals as input. The outer 25\% of phase-encode lines (32 lines at each edge of k-space) were treated as EMI-only training data. the EMI-free training targets were taken from the baseline acquisition (baseline scenario) This favors the ML methods, since the training targets share the anatomy, contrast, and scanner of the images being reconstructed, and represents a best-case scenario for the learned methods. Training inputs were synthesized by adding randomly shuffled EMI-free readout lines to these EMI realizations, with the EMI-free lines serving as training targets. Each network was trained on the dataset it was used to reconstruct. As with the other implemented methods, each of the 10 temporal frames was treated as an independent dataset in order to achieve 10 independent reconstructions to compute the quantitative metrics. Both networks were trained for 60 epochs, by which point the training loss had plateaued.

For VERGE mask generation, $s_{\text{TV}}$ was smoothed with a Gaussian kernel ($\sigma = 8$ pixels) and thresholded using Otsu's method\cite{otsu1979}. The resulting mask was cleaned with binary
morphological closing and opening, dilated with 10 iterations of a $3\times3$ cross-shaped kernel, and inverted to produce $\Omega_{\text{air}}$.

A table of parameters for both neural net implementations can be seen in table \ref{tbl:params}.

Coil combination for displaying all EMI-cleaned and baseline images was performed using BART \cite{bart}. Coil combination for the quantitative metrics was performed using SoS.

The reconstruction time for the proposed method and comparison method can be seen in \ref{tbl:times}.

No noise pre-whitening was applied.

\subsection{Evaluation Metrics}

Following previous work\cite{EDITER, STRIDE}, EMI removal performance on the \textit{in-vivo} datasets was quantified using two voxelwise metrics, SNR and EMI removal percentage, both calculated from SoS-combined images. SNR was calculated for each voxel as the mean over the 10 temporal phases divided by the standard deviation over the same phases. EMI removal percentage was calculated for each voxel as the percent change in temporal standard deviation between the EMI-corrected and the EMI-contaminated images, and is displayed in dB.

\section{Results}

\begin{figure*}[h]
  \centering
  \includegraphics[width=\textwidth]{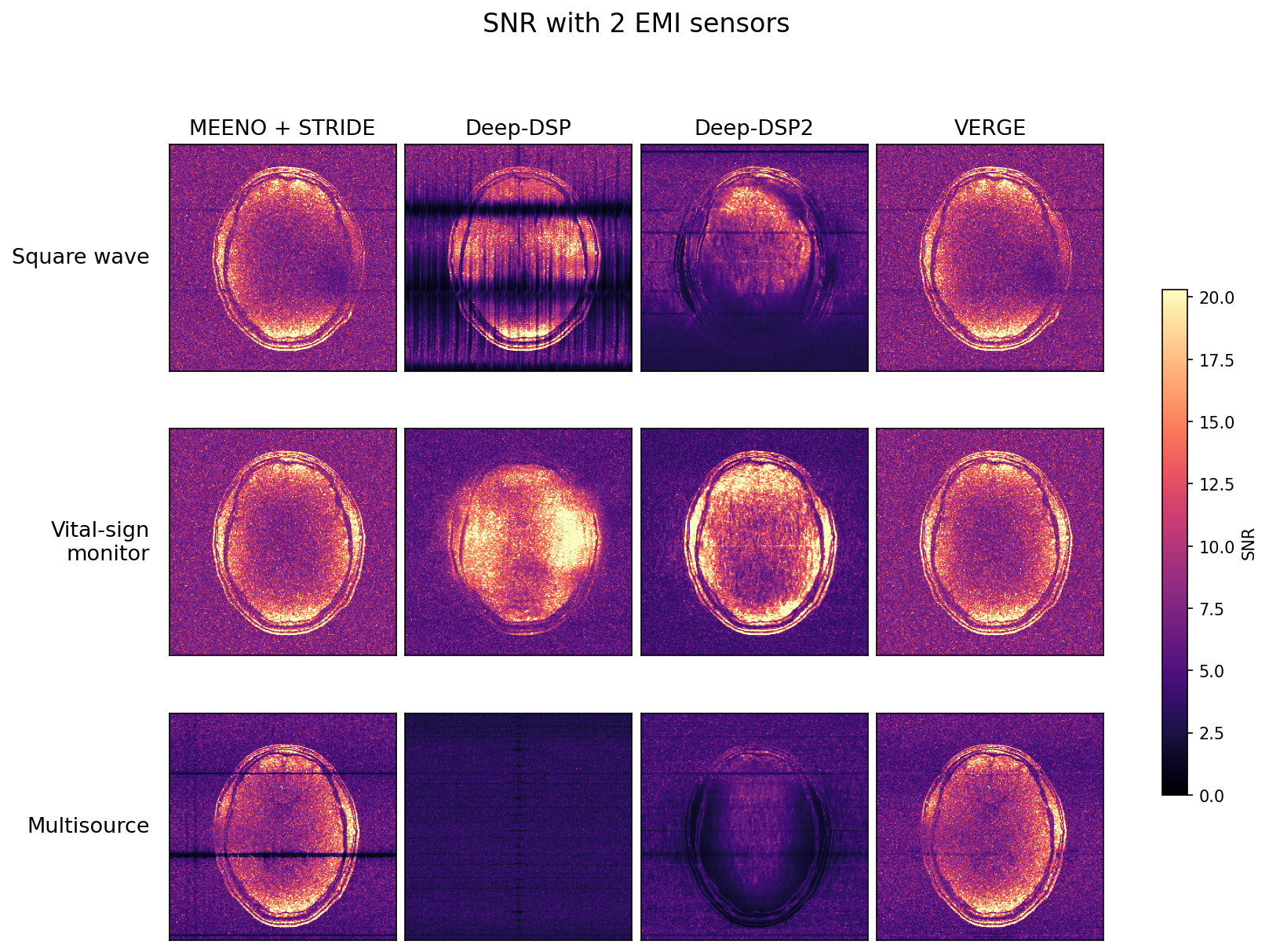}
  \caption{Visual comparison of voxel-wise SNR for each EMI removal method for the \textit{in-vivo} brain dataset when two physical EMI sensors are used for EMI removal. A higher SNR is better. VERGE is the proposed method. MEENO + STRIDE is a combination of prior methods found to outperform each individual prior method \cite{STRIDE}. Deep-DSP and Deep-DSP$^2$ are two recent ML based removal methods. VERGE non-linearly learns a virtual coil that is used in conjunction with STRIDE to remove EMI, whereas MEENO + STRIDE use strictly linear methods to remove EMI \cite{STRIDE}. This figure shoes that VERGE and MEENO + STRIDE have similar performance on the square wave and vital-sign monitor scenarios, and that VERGE outperforms MEENO + STRIDE in the multisource scenario, specifically in regions of narrow-band EMI. Deep-DSP and Deep-DSP$^2$ have low SNR in the square wave and multisource scenarios, and high SNR in the vital-sign monitor scenario.}
  \label{fig:2coilSNR}
\end{figure*}

\begin{figure*}[h]
  \centering
  \includegraphics[width=\textwidth]{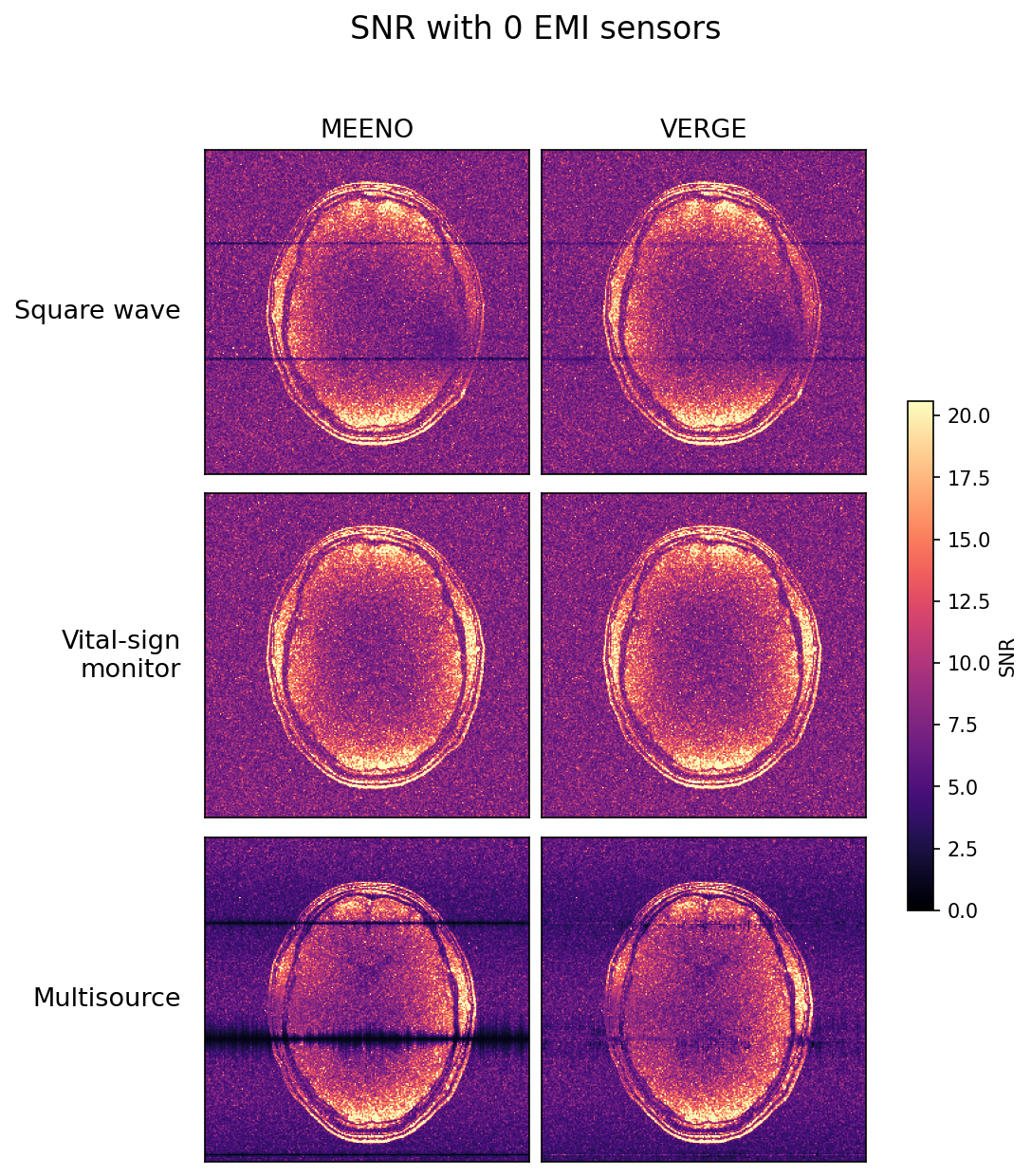}
  \caption{Visual comparison of voxel-wise SNR for each EMI removal method for the \textit{in-vivo} brain dataset when zero physical EMI sensors are used for EMI removal. A higher SNR is better. VERGE is the proposed method. MEENO + STRIDE is a combination of prior methods found to outperform each individual prior method \cite{STRIDE}. Similar to the case of two physical coils in figure \ref{fig:2coilSNR}, VERGE non-linearly learns a virtual coil that is used in conjunction with STRIDE to remove EMI, whereas MEENO use strictly linear methods to remove EMI \cite{STRIDE}. This figure shows that although the SNR is similar for the square wave and vital-sign monitor EMI scenarios, VERGE results in improved SNR for the multisource dynamic EMI from combining the two individual sources. This is specifically visible in areas corrupted by narrow-band EMI.}
  \label{fig:0coilSNR}
\end{figure*}

The SNR metrics for the \textit{in-vivo} dataset can be seen in figures \ref{fig:2coilSNR} and \ref{fig:0coilSNR}. Figure \ref{fig:2coilSNR} shows equal SNR between VERGE, STRIDE + MEENO, and the two ML methods for the single-source square wave and vital-sign monitor EMI scenarios when two physical EMI sensors are used. VERGE has a similar SNR to MEENO + STRIDE in the square wave and vital sign monitor cases, and a higher SNR in the multisource case, especially in the narrow-band EMI regions. Both ML methods have poor SNR in the square wave and multisource cases, and high SNR in the vital-sign monitor case. Figure \ref{fig:0coilSNR} shows the same metric for the scenario where zero physical EMI sensors are used. Similarly, VERGE results in improved EMI over MEENO in the case of multisource EMI corruption.

\begin{figure*}[h]
  \centering
  \includegraphics[width=\textwidth]{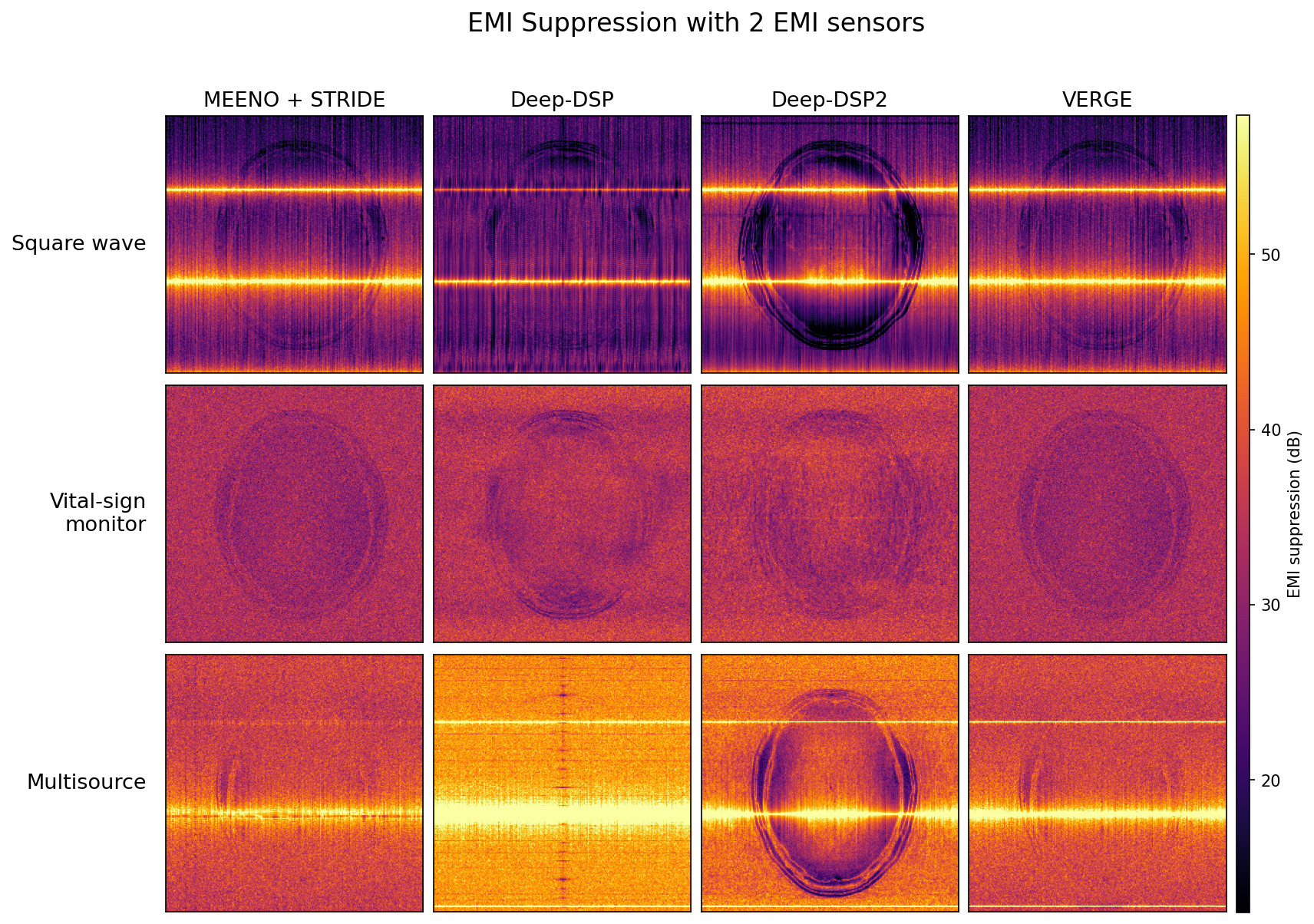}
  \caption{Visual comparison of the voxel-wise EMI removal percentage for each EMI removal method and EMI scenario in the \textit{in-vivo} dataset presented in dB in the case of two physical sensors used for EMI removal. VERGE Is the proposed method. For the square wave and vital-sign monitor scenarios, the EMI removal is similar for both MEENO + STRIDE and VERGE. In the scenario of multisource dynamic EMI corruption, VERGE has visually better EMI removal percentage in the areas of narrow-band corruption, indicating an improvement over the comparison methods. Deep-DSP and Deep-DSP$^2$ have reasonable EMI removal in the vital-sign monitor compared to the other methods and low EMI removal in the square wave scenario. In the multisource scenario, Deep-DSP has high EMI removal and Deep-DSP$^2$ has poor EMI removal in the brain.}
  \label{fig:2coilremoval}
\end{figure*}

\begin{figure*}[h]
  \centering
  \includegraphics[width=\textwidth]{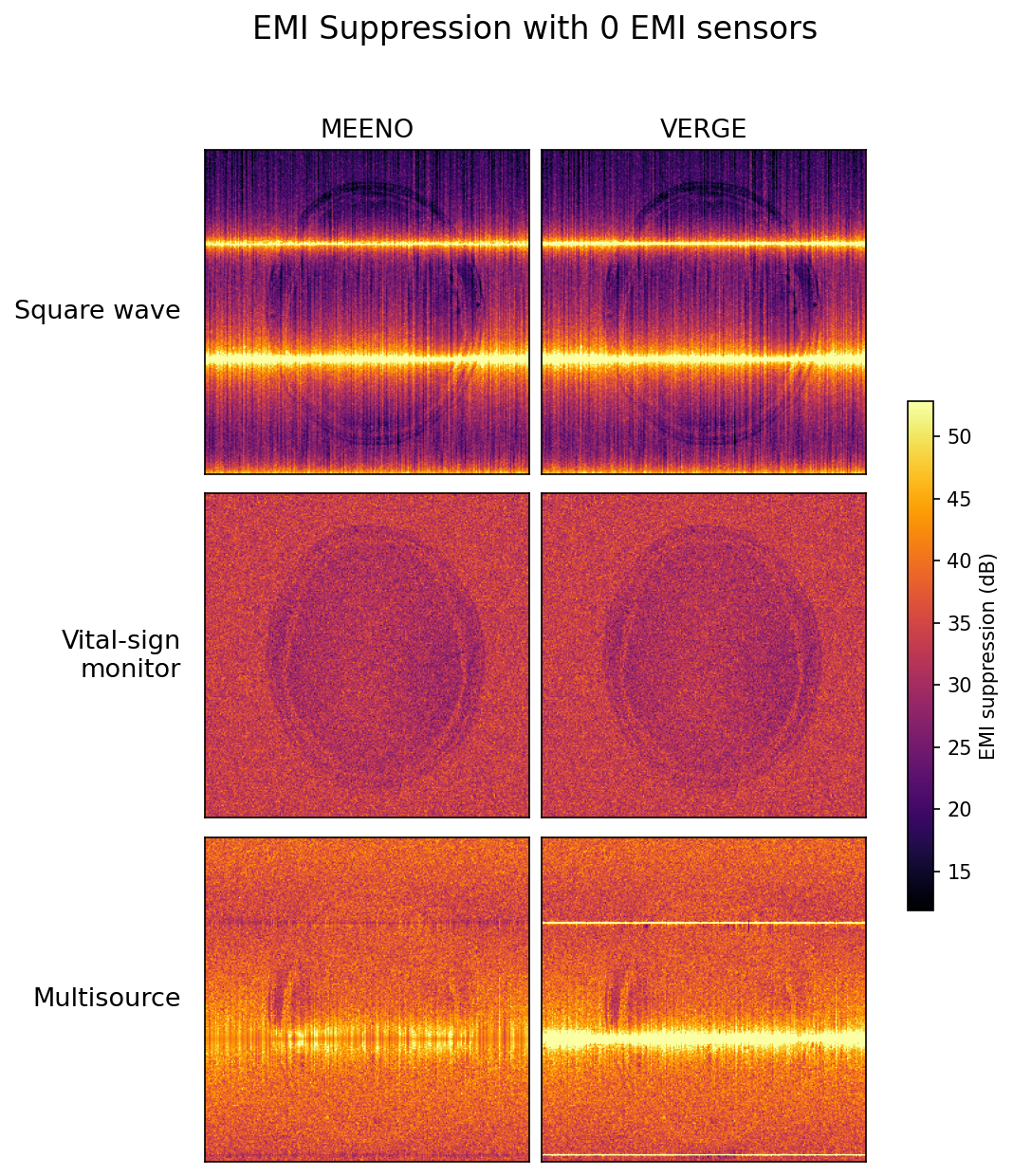}
  \caption{Visual comparison of the voxel-wise EMI removal percentage for each EMI removal method and EMI scenario in the \textit{in-vivo} dataset presented in dB in the case of zero physical sensors used for EMI removal. VERGE Is the proposed method. Similar to what is shown in figure \ref{fig:2coilremoval}, the EMI removal is similar for both MEENO and VERGE in the cases of square wave and vital-sign monitor EMI corruption. In the scenario of multisource dynamic EMI corruption, VERGE has visually better EMI removal percentage in the areas of narrow-band corruption, indicating an improvement over the comparison methods.}
  \label{fig:0coilremoval}
\end{figure*}

Similarly, EMI removal maps presented in dB can be seen in figures \ref{fig:2coilremoval} and \ref{fig:0coilremoval}. Figure \ref{fig:2coilremoval} shows the EMI removal percentage in dB between VERGE and MEENO + STRIDE for the single-source square wave and vital-sign monitor EMI scenarios when two physical EMI sensors are used. Similar to what is seen in the SNR metrics, VERGE has better EMI removal than MEENO + STRIDE in the narrow-band region for the multisource case, and is similar to MEENO + STRIDE otherwise. Deep-DSP has relatively worse removal percentage in the square wave case, similar removal percentage in the vital-sign monitor case, and relatively high removal percentage in the multisource case. Deep-DSP$^2$ has good EMI removal percentage in the square wave and vital-sign monitor case, and relatively poor removal percentage in the brain for the multisource case. Figure \ref{fig:0coilremoval} shows the same metric when zero physical EMI sensors are used. Here, the improvement with VERGE is more pronounced over MEENO + STRIDE.

\begin{figure*}[h]
  \centering
  \includegraphics[width=\textwidth]{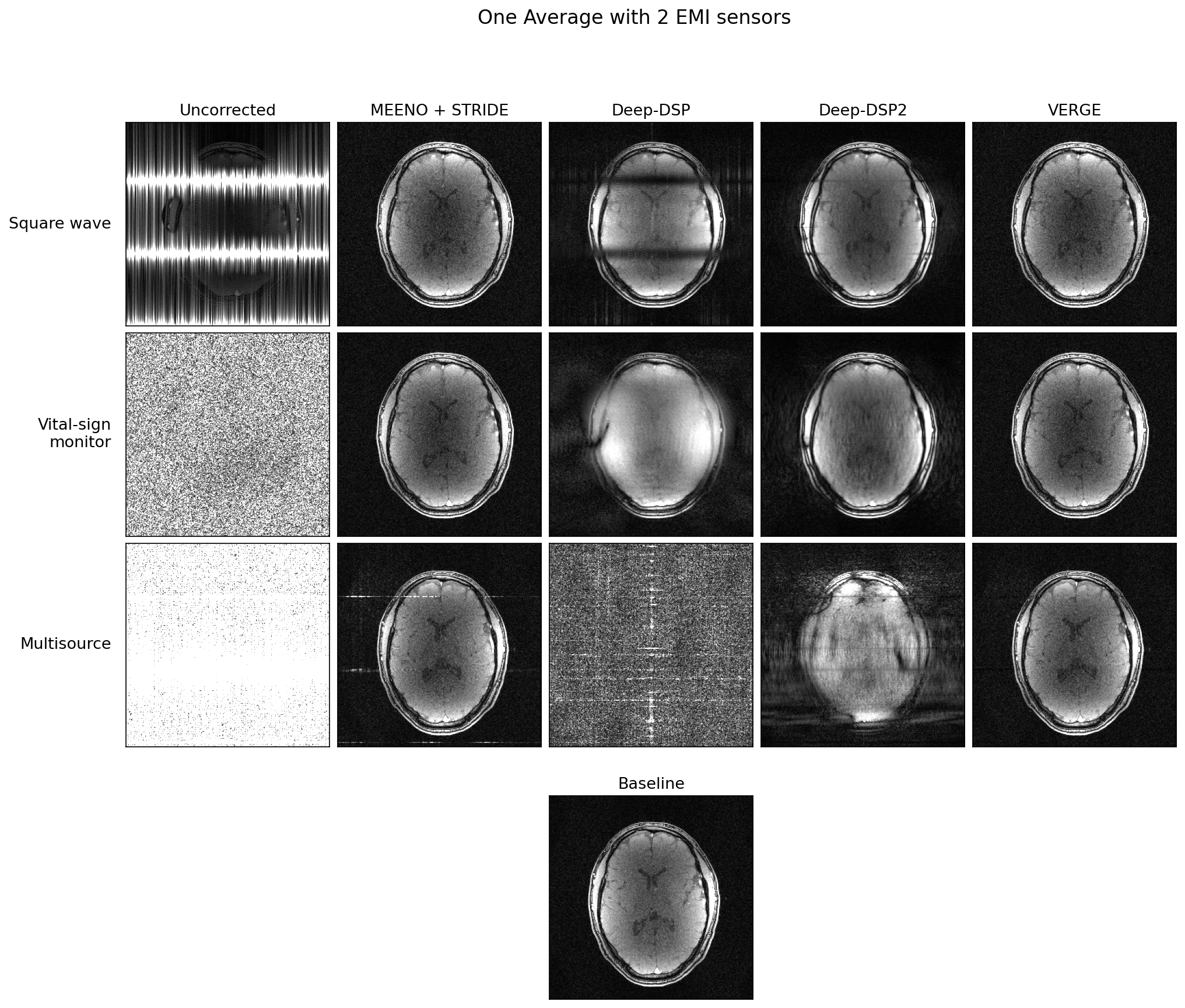}
  \caption{Visual comparison of (first column) EMI corrupted images with no EMI removal, (second column) EMI corrected images using MEENO + STRIDE, (third column) EMI corrected images using Deep-DSP, (fourth column) EMI corrrected images using Deep-DSP$^2$, and (fifth column) EMI corrected images using VERGE in the \textit{in-vivo} brain dataset using two physical EMI sensors. One single average is shown. In the case of multisource EMI, VERGE better removes the narrow-band EMI than MEENO + STRIDE. A single average baseline image is also shown.}
  \label{fig:2coil1average}
\end{figure*}

\begin{figure*}[h]
  \centering
  \includegraphics[width=\textwidth]{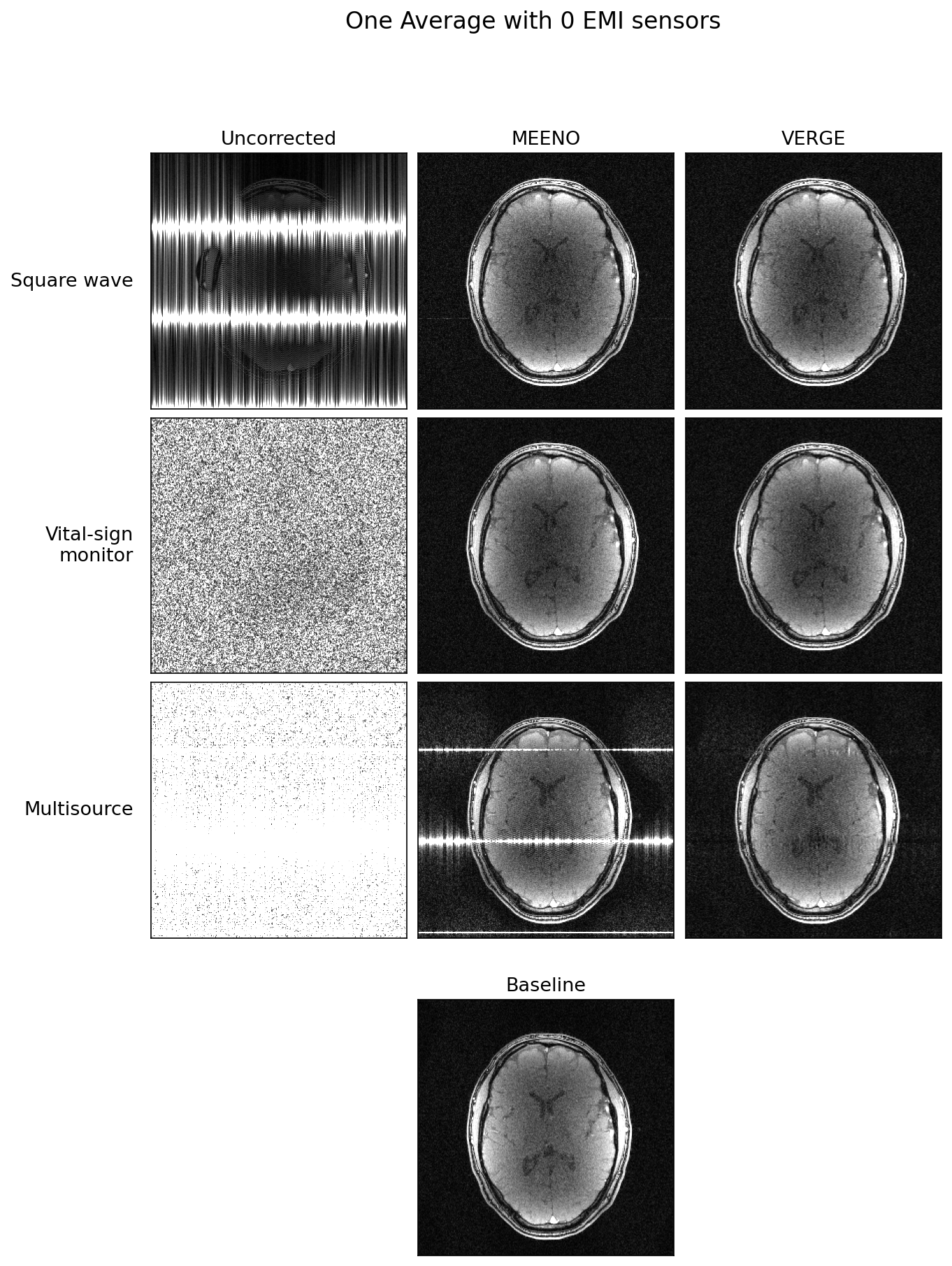}
  \caption{Visual comparison of (first column) EMI corrupted images with no EMI removal, (second column) EMI corrected images using MEENO, and (third column) EMI corrected images using VERGE in the \textit{in-vivo} brain dataset using zero physical EMI sensors. One single average is shown. Similar to what is shown in figure \ref{fig:2coil1average}, in the case of multisource EMI, VERGE better removes the narrow-band EMI than STRIDE. A single average baseline image is also shown.}
  \label{fig:0coil1average}
\end{figure*}

\begin{figure*}[h]
  \centering
  \includegraphics[width=\textwidth]{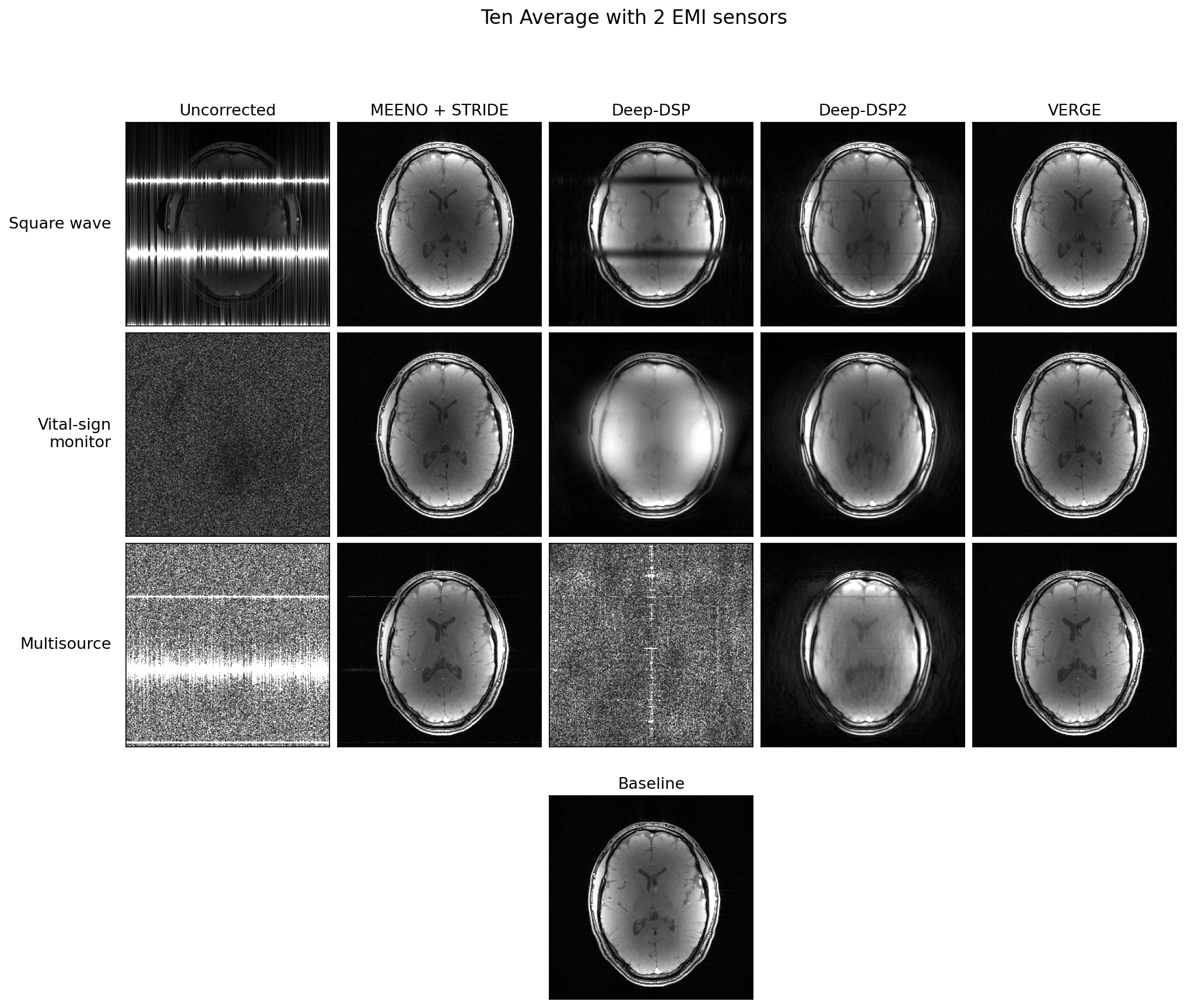}
  \caption{Visual comparison of (first column) EMI corrupted images with no EMI removal, (second column) EMI corrected images using MEENO + STRIDE, and (third column) EMI corrected images using VERGE in the \textit{in-vivo} brain dataset using two physical EMI sensors. Ten averages were used to compute the final image. In the case of multisource EMI, VERGE better removes the narrow-band EMI than MEENO + STRIDE even with ten averages. A ten average baseline image is also shown.}
  \label{fig:2coil10average}
\end{figure*}

Image reconstruction comparisons for the \textit{in-vivo} in figures \ref{fig:2coil1average} through \ref{fig:0coil10average}. Figures \ref{fig:2coil1average} and \ref{fig:0coil1average} show reconstructions with VERGE vs the comparison methods when two and zero physical EMI sensors are used respectively, and only one average is used. VERGE and MEENO + STRIDE perform similarly well in the square wave and vital-sign monitor cases, and VERGE better removes the narrow-band EMI than MEENO + STRIDE in the multisource case in both figure \ref{fig:2coil1average}. When zero physical sensors are used in figure \ref{fig:0coil1average}, MEENO leaves significant narrow-band interference whereas VERGE only has slight blurring in those areas. Deep-DSP in figure \ref{fig:2coil1average} over-removes narrow-band EMI in the square wave case, and does not produce good reconstructions in the remaining two cases. In the same figure, Deep-DSP$^2$ slightly over-removes narrow-band EMI in the square wave case, produces a slightly blurry reconstruction in the vital-sign monitor case, and produces a poor reconstruction in the multisource case.

Figures \ref{fig:2coil10average} and \ref{fig:0coil10average} show reconstructions with VERGE vs the comparison method when two and zero physical EMI sensors are used respectively, and ten averages are used. Similar to before, the square wave and vital-sign EMI scenarios are well cleaned by both deterministic methods in both figures. In the multisource scenario when two physical sensors are used, VERGE effectively eliminates EMI contamination, whereas MEENO leaves slight streaking where narrow-band EMI corruption occurred. Both ML methods over-remove EMI in regions of narrow-band interference in the square wave EMI scenario, especially with Deep-DSP. We also notice subtle dark bands in both methods that are more pronounced in Deep-DSP$^2$ that appear to be in locations where there is no narrow-band interference. This is addressed in the discussion section. Deep-DSP2 produces poor reconstructions for the remaining two scenarios, with no recognizable brain in the multi-source scenario. Deep-DSP$^2$ produces blurry reconstructions in the remaining two scenarios.

\section{Discussion}

We have shown that using a non-linearly estimated virtual EMI sensing coil improves the performance of reference-coil-based EMI removal methods. In the case of multisource dynamic EMI contamination, VERGE resulted in better SNR and EMI removal percentage than MEENO and MEENO + STRIDE, and produced visually cleaner images. Even in the case of ten averages, VERGE produced visibly cleaner images whereas the competing methods were unable to eliminate all of the EMI. VERGE with zero physical sensors outperformed MEENO + STRIDE with two.

We believe that this improvement comes from estimating the virtual coil non-linearly. The comparison methods in this work consist entirely of linear operations, meaning their ability to separate EMI from signal is limited by their ability to describe the respective subspaces. With our approach, the virtual coil is a non-linear function of the acquired data, so it can express EMI that lies outside of the subspace computed by linear methods. We acknowledge that we have not specifically tested this hypothesis. This could be tested by using the same network without activation functions.

Although we use a CNN to minimize our objective function, no training data are required. The CNN acts as a non-convex optimization solver that provides implicit spatial regularization of the virtual coil through its architecture, similar to the deep image prior and implicit neural representations\cite{ulyanov2018dip, vavasour2026inr}. Directly solving for the virtual coil with standard non-convex optimization tools did not result in satisfactory EMI removal, potentially due to the underdetermined nature of the problem.

Unlike conventional machine learning methods, we do not collect training data through pulse sequence modification or otherwise, have no train/test split, and do not attempt to generalize the network beyond a single acquisition. We therefore compare VERGE to analytical EMI removal methods rather than to learning-based approaches.

A limitation of the presented method is the additional compute time required for reconstruction. The comparison methods were able to perform EMI removal in under a second, but VERGE took minutes to reconstruct a slice. The implementation of VERGE was not done to optimize reconstruction time, but to optimize reconstruction quality. Although we may not be able to reduce EMI removal time to be similar to the comparison methods, we believe it can be significantly reduced through better hardware and algorithmic considerations. Examples of methods to lower compute time are GPU acceleration, parallelization, or fewer iterations.

The constituent methods of this work operate in both the image domain and in k-space, but no individual step requires that the sampling be Cartesian. Therefore, VERGE is in principle extendable to non-Cartesian sampling patterns; however, non-Cartesian trajectories produce different EMI artifacts in the image domain and the claim that VERGE can operate on non-Cartesian datasets must therefore be experimentally demonstrated. In future work, we will explore the effect of sampling pattern on the EMI removal of VERGE.

We have found it to be important that the mask does not contain any anatomy. Even when only a few voxels of anatomy are present in the mask, the EMI removal is insufficient and the reconstructed image is contaminated with artifacts. This is because the neural net will treat these areas as background to be suppressed. The current mask generation algorithm succeeded in generating a mask in every case, but previous iterations occasionally failed which lead us to the above conclusions.

VERGE is also dependent on an initial round of EMI removal through MEENO. Although this initial round is not meant to clean the image entirely of EMI, it is assumed that the EMI removed through MEENO does not contain anatomy. This is important because the second term in the neural net objective function assumes that the removed EMI is only EMI. Although no anatomical structure was visible in $x_{emi}$ for any dataset, it is in theory possible for anatomy to leak into the EMI through a poor estimation of the EMI subspace.

We believe that the reason for Deep-DSP and Deep-DSP$^2$ performing poorly on our data is primarily due to the number of external EMI sensors we have. In their original publications, the authors use ten external sensors whereas we only have two. We found that using a larger percentage of k-space for training produced less blurry results, but corrupted the image in other ways likely due to actual brain signal becoming present in the training data. Using a dedicated EMI characterization window in addition to the k-space periphery for training may be able to significantly improve the reconstructions since there would be more training data available to use.

We also notice that in the square wave scenario, both ML methods produce extra dark bands reminiscent of narrow-band over-removal even where there was no narrow-band interference. It is more pronounced in Deep-DSP$^2$ and nearly invisible in the single-average figure. Since this is visible in Deep-DSP, which has code available on github that was used to reproduce the method, we do not believe this to be an implementation issue with either of the ML methods. Instead, we hypothesize that it results from both networks representing the complex MR signal as separate real and imaginary channels processed with real-valued convolutions. Although cross-channel mixing allows these networks to treat positive and negative frequencies differently, nothing in the architecture enforces this. The limited training data may instead cause them to learn learn approximately conjugate-symmetric filtering, suppressing the narrow-band interference at both its actual frequency and the mirrored frequency. The additional signal voids appearing at locations mirrored about the image center from the true interference bands support this hypothesis. Although further testing could validate or invalidate our hypothesis, we believe this to be outside the scope of this paper.

Lastly, despite having high SNR and comparable EMI removal percentage in the vital-sign monitor scenario, Deep-DSP and Deep-DSP$^2$ do not produce good reconstructions. This shows that any one metric is not a good indicator of performance on its own, and must be used in conjunction with other evaluation methods.

\section{Conclusion}
We have successfully developed an EMI elimination technique that improves over the latest reference-coil-based techniques. By non-linearly generating a virtual EMI reference coil, we are able to supplement existing reference coil methods and effectively remove dynamic multisource EMI with higher SNR, EMI removal percentage (dB), and visual quality.

\bigskip
\section*{Acknowledgments}
Thank you to Gilbert Thevathasan for constructing the EMI detection coils, and to Tak Fan for helping construct the casing for the EMI detection coils. We would also like to acknowledge the support of CRANIA, and Drs Alexandre Boutet and Taufik Valiante for their support.

\subsection*{Financial disclosure}

The authors acknowledge support from the Ontario Research Fund - Grant: ORF-RE012-009 and INOVAIT - Grant: 2022-4056.

\subsection*{Conflict of interest}

The authors declare no potential conflict of interests.

\bibliographystyle{plain}
\bibliography{references}

\bigskip
\begin{table}[htbp]
\centering
\setlength{\tabcolsep}{1pt}
\caption{Neural network parameters, identical for mask generation and virtual noise coil generation unless noted.}
\label{tab:nn_params}
\begin{tabular}{ll}
\toprule
\textbf{Parameter} & \textbf{Value} \\
\midrule
\multicolumn{2}{l}{\textit{Architecture}} \\
Convolutional layers & 4 \\
Kernel size & $3\times3$ \\
Channels per layer & 64 \\
Dilation factors & 1, 2, 4, 1 \\
Receptive field & $17\times17$ pixels \\
Activation & ReLU \\
Pooling & None \\
Input channels & 32 (zero reference coils) \\
               & 36 (two reference coils) \\
Input normalization & Unit standard deviation \\
Output layer & $1\times1$ convolution \\
                & 2 channels \\
Output layer initialization & $\mathcal{N}(0,\,10^{-6})$ weights\\
& Zero bias \\
Output scaling & $\operatorname{median}(\lvert x_{\text{raw}}\rvert)$ \\
\midrule
\multicolumn{2}{l}{\textit{Training}} \\
Optimizer & Adam \\
Learning rate & $2\times10^{-3}$\\
&cosine annealed to $10^{-4}$ \\
Iterations & 500 \\
Batch size & 1 \\
Random seed & 0 \\
\midrule
\multicolumn{2}{l}{\textit{Objective}} \\
Mask generation & Normalized TV  \\
Virtual noise coil generation & $\mathcal{L}_1 + \gamma\,\mathcal{L}_2$, $\gamma = 0.3$  \\
\bottomrule
\end{tabular}\label{tbl:params}
\end{table}

\begin{table}[htbp]
\centering
\setlength{\tabcolsep}{4pt}
\caption{Reconstruction times per frame for a single slice including training time where applicable ($512\times256$ matrix, 16 imaging coils).
Times are measured on an Apple M4 Pro CPU using PyTorch 2.8.0 without GPU acceleration. Network training was not optimized for reconstruction time, but for the best EMI removal result.}
\label{tab:runtime}
\vspace{2pt}
\begin{tabular}{lcc}
\toprule
\textbf{Step} & \textbf{0 sensors} & \textbf{2 sensors} \\
\midrule
MEENO & 0.33~s & -- \\
MEENO + STRIDE & -- & 0.39~s \\
Deep-DSP & -- & 60~s \\
Deep-DSP2 & -- & 630~s \\
VERGE & 492~s & 534~s \\
\bottomrule
\end{tabular}\label{tbl:times}
\end{table}

\bigskip
\bigskip

\bigskip

\bigskip

\bigskip

\bigskip

\bigskip

\bigskip
\begin{figure*}[h]
  \centering
  \includegraphics[width=\textwidth]{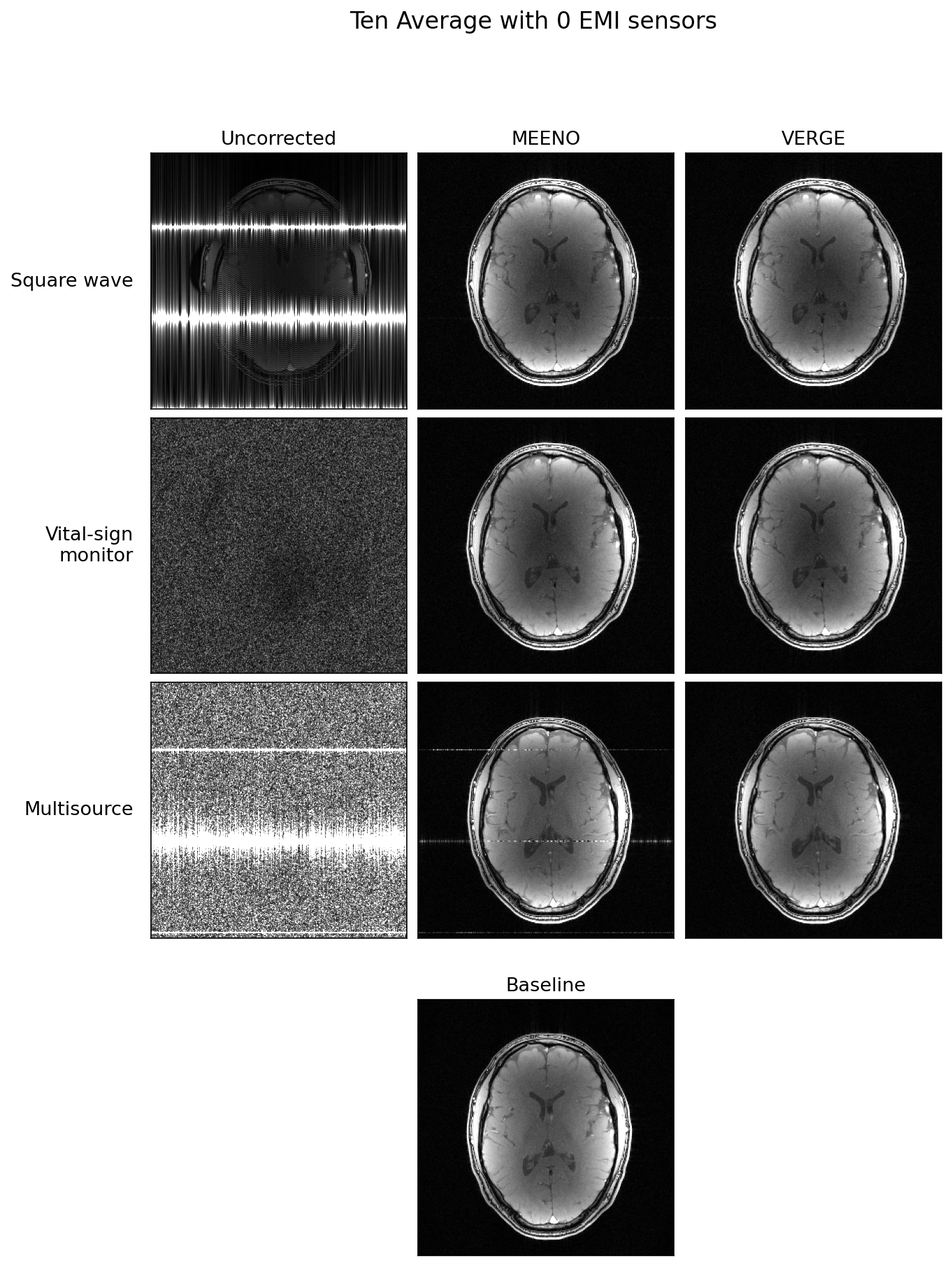}
  \caption{Visual comparison of (first column) EMI corrupted images with no EMI removal, (second column) EMI corrected images using MEENO, and (third column) EMI corrected images using VERGE in the \textit{in-vivo} brain dataset using zero physical EMI sensors. Ten averages were used to compute the final image. Similar to what is shown in figure \ref{fig:2coil10average}, in the case of multisource EMI, VERGE better removes the narrow-band EMI than STRIDE even with ten averages. A ten average baseline image is also shown.}
  \label{fig:0coil10average}
\end{figure*}

\bigskip

\section*{Supporting information}
\begin{figure*}[h!]
  \setcounter{figure}{0}
  % redefine numbering just for this float:
  \renewcommand{\thefigure}{S\arabic{figure}\ }
  \captionsetup{ }
  \centering
  \includegraphics[width=\textwidth]{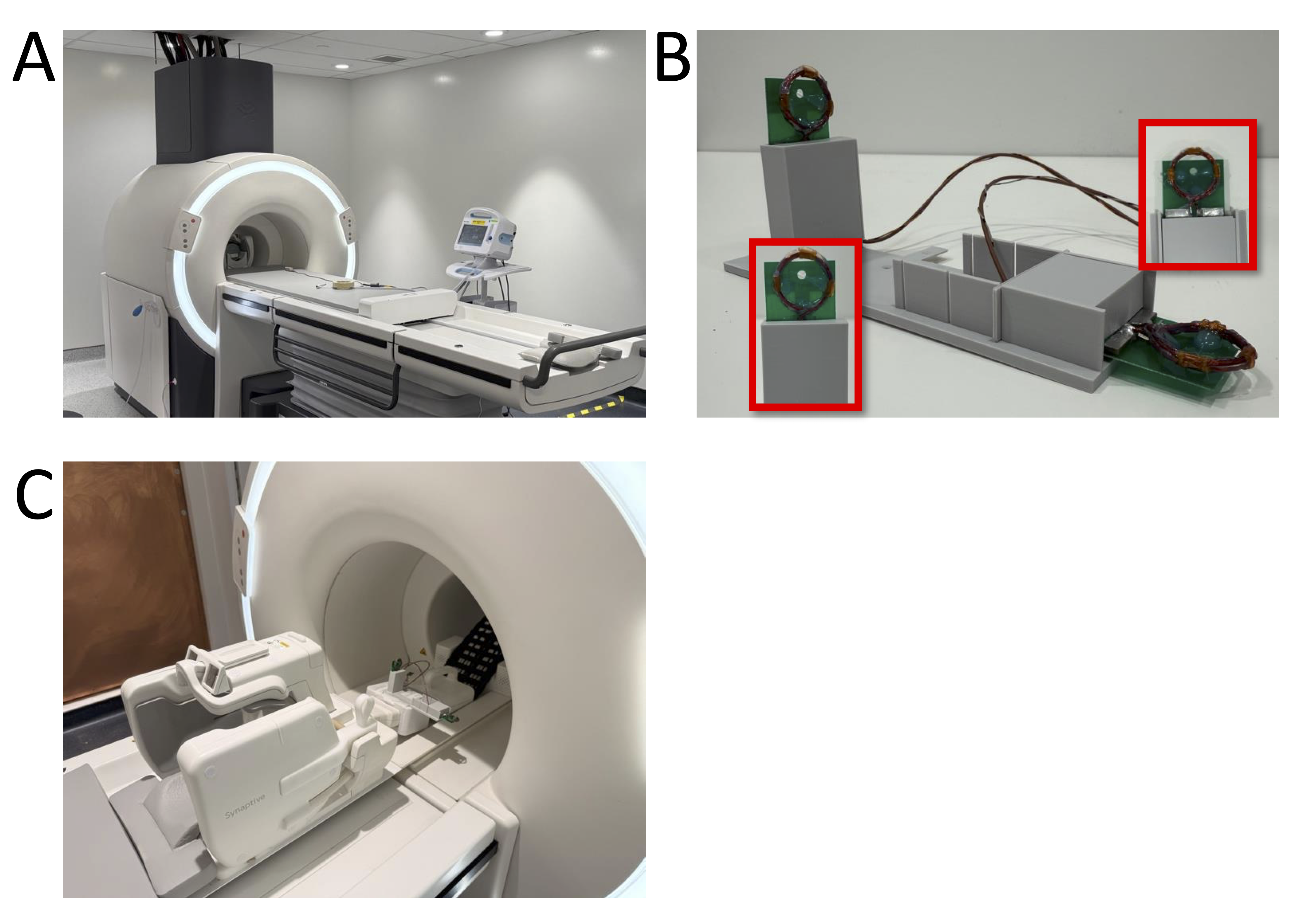}
  \caption{Setup of the WA machine for measuring EMI (A), and the EMI sensing coils (B), and the coils' position relative to the head coil while scanning (C)}
  \label{sup:figure_s1}
\end{figure*}

\end{document}